\documentclass[12]{article}
\usepackage{fleqn}
\DeclareMathAlphabet{\mathpzc}{OT1}{pzc}{m}{it}
\usepackage{amsmath}
\usepackage{graphicx}
\begin{document}

\Large
\begin{center}
\bf{A Finite-Geometric Classification of Three-Qubit Mermin Pentagrams}
\end{center}
\vspace*{-.3cm}

\large
\begin{center}
 Metod Saniga$^{1}$, Fr\'ed\'eric Holweck$^{2}$ and Hamza Jaffali$^{3}$
\end{center}
\vspace*{-.3cm} 

\normalsize
\begin{center}

$^{1}$Astronomical Institute of the Slovak Academy of Sciences,\\
SK-05960 Tatransk\' a Lomnica, Slovak Republic\\
(msaniga@astro.sk)
\vspace*{0.1cm}

$^{2}$Laboratoire Interdisciplinaire Carnot de Bourgogne, ICB/UTBM, UMR 6303 CNRS,
Universit\'e Bourgogne Franche-Comt\'e, F-90010 Belfort, France\\ (frederic.holweck@utbm.fr)
\vspace*{0.0cm}

and
\vspace*{0.0cm}

$^{3}$UTBM, Universit\'e Bourgogne Franche-Comt\'e, F-90010 Belfort, France\\ (hamza.jaffali@utbm.fr)

\end{center}

\vspace*{-.4cm} \noindent \hrulefill

\vspace*{-.1cm} \noindent {\bf Abstract}

\noindent
Given the facts that the three-qubit symplectic polar space features three different kinds of observables and each of its labeled Fano planes acquires a definite sign, we found that there are 45 distinct types of Mermin pentagrams in this space. A key element of our classification is the fact that any context of such pentagram is associated with a unique (positive or negative) Fano plane.
Several intriguing relations between the character of pentagrams' three-qubit observables and `valuedness' of associated Fano planes are pointed out. In particular, we find two distinct kinds of negative contexts and as many as four positive ones.
\\

\vspace*{-.2cm}

%\vspace*{-.2cm}
%\noindent
%{\bf MSC:} 05B25 -- 51E20 

\vspace*{-.2cm} \noindent \hrulefill

%\bigskip
\noindent
\section{Introduction}
A Mermin pentagram, first introduced by Mermin in 1993 \cite{mer} to furnish an observable-based proof of quantum contextuality \cite{ks}, is a set of ten observables of a three-qubit system with eigenvalues $\pm 1$ that are arranged into five four-element sets (contexts) that lie along the five edges of a pentagram in such a way that four observables in the same set mutually commute, their product is $+1$ or $-1$, and the number of sets where the latter holds is odd. Some years ago, using computer, it was shown \cite{psh} that the symplectic polar space associated with the three-qubit Pauli group contains as many as 12\,096 such pentagrams, forming three distinct families according as the number of negative contexts is five (108 members), three (4104 members) or one (7884 members). More recently, L\'evay and Szab\'o 
\cite{ls} analyzed this symplectic polar space in terms of its magic Veldkamp line and discovered that Mermin pentagrams form naturally aggregates of six pairs, each such double-six `cell' being intricately related to six spreads of the core doily (generalized quadrangle of order two). In this short note we shall reveal further interesting traits in the structure of Mermin pentagrams by taking into account that the elements of the three-qubit Pauli group (three-qubit observables) are of three different kinds and employing an important earlier observation \cite{salev} that each edge of a Mermin pentagram corresponds to a unique Fano plane of the associated symplectic polar space with a fixed/prescribed three-qubit labeling of its points.

\section{3-Qubit Observables and Positive/Negative Fano Planes}
The (generalized) three-qubit Pauli group, ${\cal P}_3$, is generated by three-fold tensor products of the matrices
\begin{eqnarray*}
I = \left(
\begin{array}{rr}
1 & 0 \\
0 & 1 \\
\end{array}
\right),~
X = \left(
\begin{array}{rr}
0 & 1 \\
1 & 0 \\
\end{array}
\right),~
Y = \left(
\begin{array}{rr}
0 & -i \\
i & 0 \\
\end{array}
\right)
~{\rm and}~
Z = \left(
\begin{array}{rr}
1 & 0 \\
0 & -1 \\
\end{array}
\right).
\end{eqnarray*}
Explicitly,
\begin{eqnarray*}
{\cal P}_3 = \{i^{\alpha} G_1 \otimes G_2 \otimes G_3:~ G_j \in \{I, X, Y, Z \},~ j \in \{1, 2, 3\},~\alpha \in \{0, 1, 2, 3\} \}.
\end{eqnarray*}
Here, we will only be dealing with its factored version $\overline{{\cal P}}_3 \equiv {\cal P}_3/{\cal Z}({\cal P}_3)$, where the center ${\cal Z}({\cal P}_3)$ consists of $\pm I \otimes I \otimes I$ and $\pm i I \otimes I \otimes I$,\footnote{In what follows, we shall use a shorthand notation for the tensor product: $G_1 \otimes G_2 \otimes G_3 \equiv G_1 G_2 G_3$.} and whose geometry is that of the symplectic polar space $W(5,2)$ \cite{sp07}. This space is, freely speaking, a collection of all totally isotropic subspaces of the ambient five-dimensional binary projective space, PG$(5,2)$, equipped with a non-degenerate alternating bilinear form.

\begin{figure}[t]
\centerline{\includegraphics[width=9.cm,clip=]{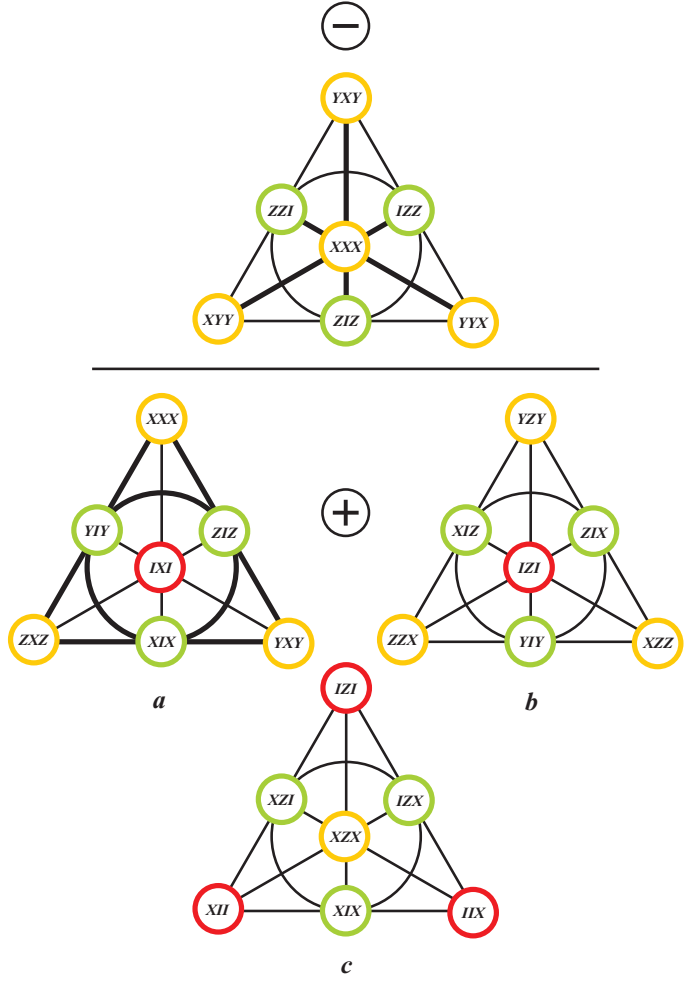}}
\caption{A representative of the family of negative Fano planes (top) and those of three distinct types of positive Fano planes. Negative lines are shown in bold. An observable of type $A$, $B$ and $C$ is colored red, green and yellow, respectively.}
\end{figure}

The 63 non-trivial elements of the group are in a bijective correspondence with the 63 points of $W(5,2)$ in such a way that two commuting elements correspond to two points joined by a totally isotropic line and a maximum set of mutually commuting elements of the group has its counterpart in a maximal totally isotropic subspace, which is a projective plane of order two, the Fano plane. Let us assume that $W(5,2)$ has its points labeled by the elements of  $\overline{{\cal P}}_3$. A line/plane of such a space is called positive or negative according as the product of the group elements located in it is  $+III$ or $-III$, respectively. Next, let us call an element of $\overline{{\cal P}}_3$ to be of type $A$, $B$ or $C$ in dependence on whether, respectively, it features two $I$'s, one $I$ or no $I$. With this last notion at hand, we will find that there are four different types of three-qubit-labeled Fano planes: one negative type and three positive ones. A negative Fano plane consists of three concurrent negative lines. If a positive Fano plane contains negative lines (type $a$), there are always four of them, forming the Pasch configuration. If a positive Fano plane is devoid of negative lines, then it has either one element of $\overline{{\cal P}}_3$ of type $A$ and three of type $C$  (type $b$), or {\it vice versa} (type $c$). The situation is schematically illustrated in Figure 1. 

\begin{figure}[t]
\centerline{\includegraphics[width=12.cm,clip=]{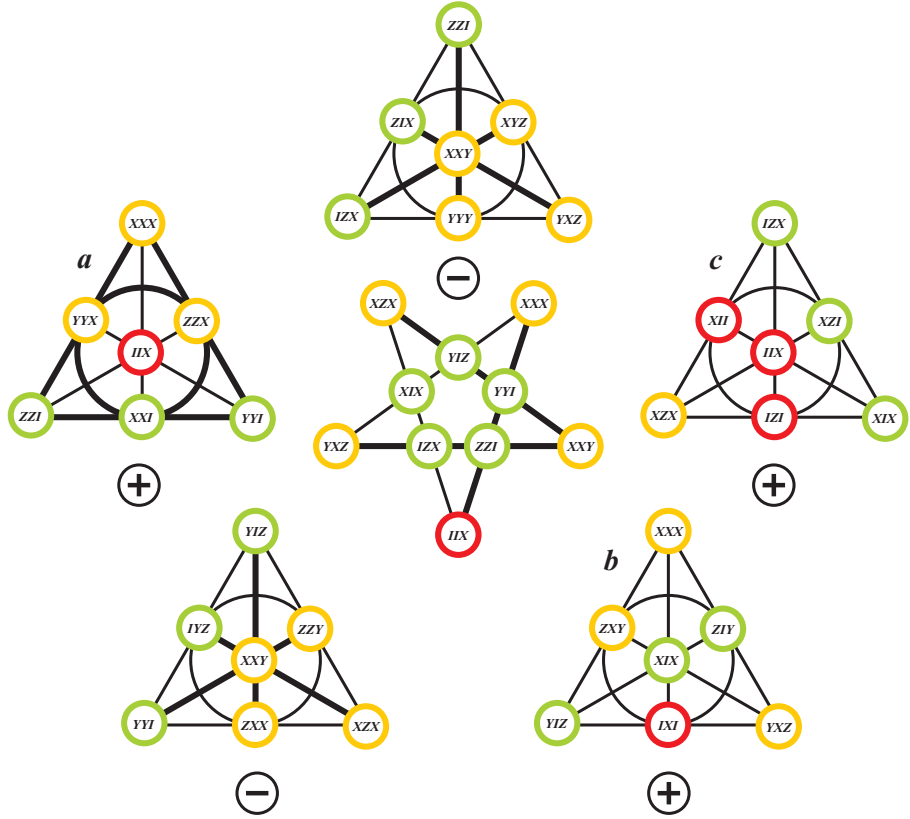}}
\caption{A Mermin pentagram (drawn in the center) and the associated pentad of labeled Fano planes. The Fano plane at the top corresponds to the horizontal edge of the pentagram; the remaining correspondences follow readily from the rotational symmetry of the figure. Both the negative contexts of the pentagram and the negative lines of three Fano planes are boldfaced. Note that one of the negative contexts corresponds to a positive Fano plane.  The Fano planes are labelled by the three-qubit observables in such a way that the closure line of the affine plane (the `line at infinity') is always represented by a (big) circle.}
\end{figure}

\begin{table}[pth!]
\begin{center}
\caption{Refined geometric classification of Mermin pentagrams. Column one ($T$) shows the type, column two ($C^{-}$) the number of negative contexts in a pentagram of the given type, columns three to five ($O_{A}$ to $O_{C}$) indicate the number of observables of corresponding types, column six ($F^{-}$) the number of negative Fano planes and columns seven to eight ($F^{+}_{a}$ to $F^{+}_{c}$) the distribution of types of positive Fano planes. Finally, the last column ($K$) indicates the number of pentagrams of a given type that lie on the `symmetric' Klein quadric. } \vspace*{0.9cm}
%\resizebox{\columnwidth}{!}{% 
\scalebox{0.8}{
\begin{tabular}{|r|c|ccc|c|ccc|r|}
\hline \hline
$T$        & $C^{-}$ & $O_{A}$ & $O_{B}$ & $O_{C}$  & $F^{-}$  & $F^{+}_{a}$  & $F^{+}_{b}$  & $F^{+}_{c}$  & $K$   \\
\hline
  1        & 5       & 0       & 0       & 10       & 5        & 0            & 0            & 0   & 2 \\
  2        & 5       & 1       & 0       &  9       & 3        & 2            & 0            & 0   & 0 \\
\hline
  3        & 3             & 0       & 5       &  5       & 4              & 0                 & 1                 & 0    & 0 \\
	4        & 3             & 0       & 4       &  6       & 5              & 0                 & 0                 & 0    & 0 \\
  5        & 3             & 0       & 4       &  6       & 3              & 2                 & 0                 & 0    & 6 \\
  6        & 3             & 0       & 4       &  6       & 3              & 1                 & 1                 & 0    & 0 \\
	7        & 3             & 0       & 4       &  6       & 3              & 0                 & 2                 & 0    & 6 \\
  8        & 3             & 1       & 0       &  9       & 3              & 0                 & 2                 & 0    & 0 \\
  9        & 3             & 1       & 4       &  5       & 3              & 2                 & 0                 & 0    & 0 \\
 10        & 3             & 1       & 4       &  5       & 3              & 1                 & 0                 & 1    & 12 \\
 11        & 3             & 1       & 4       &  5       & 3              & 0                 & 2                 & 0    & 0 \\
 12        & 3             & 1       & 4       &  5       & 3              & 0                 & 1                 & 1    & 6 \\
 13        & 3             & 1       & 4       &  5       & 3              & 0                 & 0                 & 2    & 6 \\
 14        & 3             & 1       & 5       &  4       & 2              & 2                 & 1                 & 0    & 0 \\
 15        & 3             & 1       & 5       &  4       & 2              & 1                 & 1                 & 1    & 12 \\
 16        & 3             & 2       & 5       &  3       & 2              & 1                 & 0                 & 2    & 12 \\
 17        & 3             & 2       & 4       &  4       & 1              & 2                 & 2                 & 0    & 0 \\
 18        & 3             & 2       & 4       &  4       & 1              & 2                 & 1                 & 1    & 12 \\
 19        & 3             & 3       & 0       &  7       & 1              & 2                 & 1                 & 1    & 6 \\
 20        & 3             & 3       & 4       &  3       & 1              & 2                 & 0                 & 2    & 6 \\
\hline
 21        & 1             & 0       & 5       &  5       & 4              & 1                 & 0                 & 0    & 0 \\
 22        & 1             & 0       & 5       &  5       & 2              & 2                 & 1                 & 0    & 12 \\
 23        & 1             & 0       & 5       &  5       & 2              & 1                 & 2                 & 0    & 12 \\
 24        & 1             & 0       & 4       &  6       & 3              & 1                 & 1                 & 0    & 12 \\
 25        & 1             & 1       & 4       &  5       & 3              & 0                 & 2                 & 0    & 1 \\
 26        & 1             & 1       & 4       &  5       & 3              & 0                 & 1                 & 1    & 4 \\
 27        & 1             & 1       & 4       &  5       & 3              & 0                 & 0                 & 2    & 7 \\
 28        & 1             & 1       & 4       &  5       & 1              & 2                 & 1                 & 1    & 19 \\
 29        & 1             & 1       & 4       &  5       & 1              & 0                 & 3                 & 1    & 6 \\
 30        & 1             & 1       & 5       &  4       & 2              & 2                 & 0                 & 1    & 12 \\
 31        & 1             & 1       & 5       &  4       & 2              & 1                 & 2                 & 0    & 0 \\
 32        & 1             & 1       & 5       &  4       & 2              & 1                 & 1                 & 1    & 12 \\
 33        & 1             & 1       & 5       &  4       & 2              & 0                 & 2                 & 1    & 11 \\
 34        & 1             & 2       & 5       &  3       & 2              & 1                 & 0                 & 2    & 12 \\
 35        & 1             & 2       & 5       &  3       & 2              & 0                 & 1                 & 2    & 24 \\
 36        & 1             & 2       & 4       &  4       & 1              & 2                 & 1                 & 1    & 11 \\
 37        & 1             & 2       & 4       &  4       & 1              & 1                 & 2                 & 1    & 25 \\
 38        & 1             & 3       & 0       &  7       & 1              & 0                 & 3                 & 1    & 2 \\
 39        & 1             & 3       & 4       &  3       & 1              & 1                 & 1                 & 2    & 12 \\
 40        & 1             & 3       & 4       &  3       & 1              & 0                 & 2                 & 2    & 6 \\
 41        & 1             & 3       & 5       &  2       & 0              & 2                 & 1                 & 2    & 12 \\
 42        & 1             & 3       & 5       &  2       & 0              & 1                 & 2                 & 2    & 12 \\
 43        & 1             & 4       & 4       &  2       & 1              & 1                 & 0                 & 3    & 12 \\
 44        & 1             & 4       & 4       &  2       & 1              & 0                 & 1                 & 3    & 12 \\
 45        & 1             & 6       & 0       &  4       & 1              & 0                 & 0                 & 4    & 2 \\
 \hline \hline
\end{tabular}
}
%}%
\end{center}
\end{table}

\section{Fine Structure of Mermin Pentagrams}
Given a Mermin pentagram, the first parameter of our classification is the number of negative contexts. The next three parameters are the character of three-qubit elements located on it.
By a way of example, the pentagram depicted in Figure 2 has three negative contexts and features one observable of type $A$, five of type $B$ and four ones of type $C$. Next, in Ref. \cite{salev} it was shown
that each edge of a pentagram is isomorphic to an affine plane of order two. As each such plane can uniquely be extended to the projective plane of order two, this also means that each edge is associated with the unique positive- or negative-valued Fano plane. The types of the five associated Fano planes will be the last string of parameters of our classification.  Using again the example from Figure 2, we find here two negative and three positive Fano planes, the latter falling into all three types.

In order to fully classify Mermin pentagrams in this manner, we made use of particular sets of them computed by Michel Planat \cite{plan}. One such set comprises all 336 pentagrams located on a particular Klein quadric and another set consists of those pentagrams located (as point-sets) in a particular geometric hyperplane of a copy of the split Cayley hexagon of order two embedded classically into our three-qubit symplectic polar space \cite{psh}. To check for completeness, we randomly picked up pentagrams from the whole set, for each of them we looked for those ten pentagrams each of which shares two edges with the pentagram in question and compared their types with those already found. The results of our analysis are summarized in Table 1; note that our pentagram of Figure 2 belongs to type 15.

Let us highlight some of the most interesting properties of Mermin pentagrams stemming from our analysis. 
Obviously, there are only two types where all five Fano planes are negative (types 1 and 4) or positive (types 41 and 42). No pentagram possesses less than two observables of type $C$, or has just eight of them. If a pentagram features observables of type $B$, there are four or five of them; and if an edge of such a pentagram contains one such observable, it must contain one more. Next, if a pentagram contains just three observables of type $A$, they are situated on the same edge; the only exception are types 41 and 42, since they feature no negative Fano plane. There are no pentagrams featuring more than six observables of type $A$. Any pentagram that contains positive Fano planes of type $b$ and $c$, but none of type $a$, is that with a single negative context; the only exception to this rule is type 12. One also sees that a pentagram with no observable of type $A$ is also devoid of positive Fano planes of type $c$. Further, if a pentagram is associated with Fano planes featuring all the three positive types, then this pentagram also exhibits all the three kinds of observables, with a single exception (type 19). It is also worth mentioning that there are three types of pentagrams having positive Fano planes of type $c$ only (types 13, 27 and 45), as well as two types with all positive Fano types being equally represented (types 15 and 32). 
From the physical point of view, the most important finding is certainly the existence of two different types of negative contexts of Mermin pentagrams and as many as four distinct kinds of positive ones; the former case involves negative Fano planes and positive Fano planes of type $a$,  whereas the latter one entails all the four types of valued Fano planes. A particularly nice example of these properties is furnished by a pentagram of types 28 or 36, whose single negative context is associated with a positive Fano plane and the remaining four positive contexts are all of different characters.

As a particular task, we also analyzed case by case all 336 pentagrams lying on that hyperbolic quadric (Klein quadric) of $W(5,2)$ that accommodates all 35 symmetric elements of $\overline{{\cal P}}_3$, and found out that they only fall into 34 distinct types; interestingly enough, the eleven missing types are types 2, 3, 4, 6, 8, 9, 11, 14, 17, 21 and 31, none of them being associated with a positive Fano plane of type $c$ (see the last column of Table 1).

\section{Concluding Remarks}
We have introduced a remarkable finite-geometric classification of three-qubit Mermin pentagrams that intricately combines the character of three-qubit observables with the properties of positive/negative-valued Fano planes of the associated symplectic polar space and reveals important finer structure of three-qubit (observable-based) quantum contexts, distinguishing between two negative and as many as four  positive ones.
We believe that such classification can be of relevance in any branch of quantum information theory (quantum protocols) where a Mermin pentagram is an essential element; it can also be helpful in revealing finer traits of the so-called black-hole/qubit correspondence (see, e.\,g., \cite{bdl}) in those of its aspects that are linked to the structure of the three-qubit symplectic polar space.

\newpage
\section*{Acknowledgments}
This work was supported by the Slovak Research and Development Agency under the contract $\#$ SK-FR-2017-0002 and the French Ministry of Europe and Foreign Affairs (MEAE) under the project PHC \v Stef\'anik 2018/40494ZJ. The financial support of both the Slovak VEGA Grant Agency, Project $\#$ 2/0003/16, and the French ``Investissements d'Avenir'' programme, project ISITE-BFC (contract ANR-15-IDEX-03), are gratefully acknowledged as well. We are also indebted to Dr. Petr Pracna for the help with the figures.


\vspace*{-.1cm}
\begin{thebibliography}{10}
\itemsep=-2pt
\bibitem{mer}
N. David Mermin, Hidden Variables and the Two Theorems of John Bell, Reviews of Modern Physics 65 (1993) 803.
\bibitem{ks}
S. Kochen and E. P. Specker, The Problem of Hidden Variables in Quantum Mechanics, Journal of Mathematics and Mechanics 17 (1967) 59. 
\bibitem{psh}
M. Planat, M. Saniga and F. Holweck, Distinguished Three-Qubit `Magicity' via Automorphisms of the Split Cayley Hexagon, Quantum Information Processing 12 (2013) 2535. 
\bibitem{ls}
P. L\'evay and Z. Szab\'o, Mermin Pentagrams Arising from Veldkamp Lines for Three Qubits, Journal of Physics A: Mathematical and Theoretical  50 (2017) Art. No. 095201.
\bibitem{salev}
M. Saniga and P. L\'evay, Mermin's Pentagram as an Ovoid of PG(3,2), EPL -- Europhysics Letters 97 (2012) Art. No. 50006. 
\bibitem{sp07}
M. Saniga and M. Planat, Multiple Qubits as Symplectic Polar Spaces of Order Two, Advanced Studies in Theoretical Physics 1 (2007) 1.
\bibitem{plan}
M. Planat, private communication.
\bibitem{bdl}
L. Borsten, M. Duff and P. L\'evay, The Black-Hole/Qubit Correspondence: An Up-to-Date Review, Class. Quantum Grav. 29 (2012) Art. No. 224008.
\end{thebibliography}
\end{document}